# Visible Light Assisted Gas Sensing with TiO$_2$ Nanowires

*Jie Zhang[1], Evgheni Strelcov[1*†], and Andrei Kolmakov[1**]*

[1]Department of Physics, Southern Illinois University Carbondale, Illinois, USA

**Abstract:**

Sensing response of individual single-crystal titania nanowires configured as chemiresistors for detecting reducing (CO, H$_2$) and oxidizing (O$_2$) gases is shown to be sensitive to visible light illumination. It is assumed that doping of the TiO$_2$ nanowires with C and/or N during carbon assisted vapor-solid growth creates extrinsic states in the band gap close to the valence band maximum, which enables photoactivity at the photon energies of visible light. The inherently large surface-to-volume ratio of nanowires, along with facile transport of the photo-generated carriers to/from the nanowire's surface promote the adsorption/desorption of donor/acceptor molecules, and therefore open the possibility for visible light assisted gas sensing. The photo-catalytic performance of TiO$_2$ nanowire chemiresistors demonstrates the prospect of combining light harvesting and sensing action in a single nanostructure.

**Introduction:**

Titania (TiO$_2$) is a well-known photocatalyst used in water photoelectrolysis applications for converting sunlight into a chemical fuel (hydrogen), for conversion of CO$_2$ to hydrocarbon fuels, for photodegradation of toxic molecules to simple harmless products, self-cleaning glass and fabric,[1, 2] photo-induced superwetting,[3] etc.[4, 5] However, pristine TiO$_2$ is a wide band gap semiconductor with the band gap in excess of 3 eV. Its photocatalytic activation therefore requires ultraviolet irradiation that comprises only a small fraction (~5%) of the solar constant.[6] Thus, it is desirable to develop a TiO$_2$-based photocatalyst active in the visible light (45% of the sun's radiation energy).[7] Considerable efforts have been made recently to extend the photoresponse of the TiO$_2$ systems farther into the visible-light region via compositional doping with a wide range of transition-metal cations.[8,9] These works revealed a disadvantage of the cationic dopants: they can result in localized d-electronic levels deep in the band gap of TiO$_2$, which serve as recombination centers for photogenerated charge carriers. A large part of the existing literature agrees that anionic nonmetal dopants, such as carbon,[10] sulfur,[11] and nitrogen,[12] are more appropriate for extending the TiO$_2$ photocatalytic activity into the longer wavelength region. Accordingly, the reported doping methods include chemical approaches, such as sol-gel reaction synthesis, electrochemical doping and oxidation of titanium nitride, as well as physical methods such as magnetron co-sputtering and ion implantation.[9,13,14]

[*]Present address: Center for Nanophase Materials Sciences, Oak Ridge National Laboratory, Oak Ridge, Tennessee
[**] Present address: Center for Nanoscale Science and Technology, NIST, Gaithersburg, MD
[†]Corresponding author: archerstrl@yahoo.co.uk

Historically, photo-assisted metal oxide gas sensors have been known for a long time.[15-17] Pristine TiO$_2$ nanowires (NWs), having a high surface-to-volume ratio and photocatalytically active in the UV region, are one of the most studied platforms for detection of reducing gases. In this report, TiO$_2$ nano- and meso-wires responsive to *visible light* were synthesized via vapor-solid (VS) technique with unintentional carbon and/or nitrogen doping during the carbothermal evaporation.

**Experimental**

N-type TiO$_2$ nanowires 50-1000 nm wide and hundred microns long were synthesized via vapor-solid carbothermal method using a slightly modified protocol described previously.[18-20] Namely, a precursor mixture of TiO$_2$ and graphite powders (1:1 by volume) was heated in a tube furnace at 1100 °C in an Ar carrier gas atmosphere (30 sccm, 200 Torr) containing traces of air at $10^{-2}$ Torr. Pure Ti (99.99%) polycrystalline samples placed downstream in the 750~800 °C zone served as collector substrates. Ti surface becomes oxidized by residual oxygen during the temperature ramp. Newly formed micron thick polycrystalline TiO$_2$ skin provides plenty of nucleation sites for preferential 1D titania NW growth (Fig 1a). The morphology and crystalline structure of the grown TiO$_2$ NWs were characterized by Scanning Electron Microscopy (SEM) and X-Ray Diffraction analysis (XRD), as shown in Figure 1. The nanowires collected from the substrates were placed on the oxide side of a Si/(300 nm)-SiO$_2$ wafer. Ti (20 nm)/Au (200 nm) micro-pad contacts were thermal-vapor deposited through a shadow mask (Fig. 2 inset). The Si substrate (p-doped) back gate electrode was grounded during the measurements.

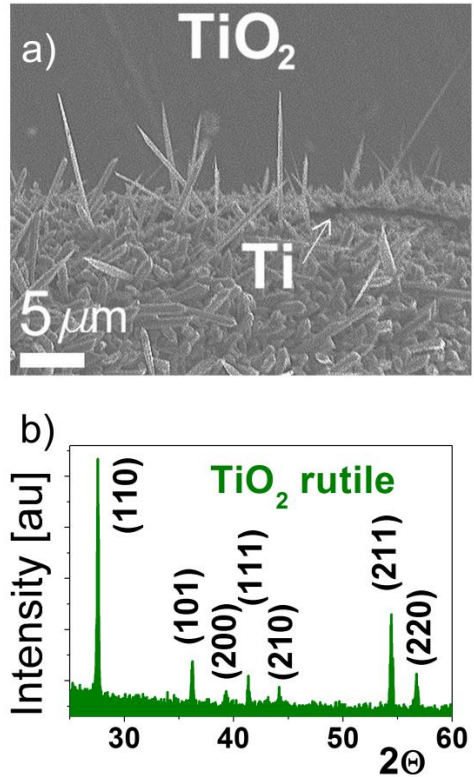

**Figure 1.** a) SEM micrograph of the synthesized $TiO_2$ nanostructures; b) XRD pattern of nanowires reveals their rutile structure.

The photocatalytic and gas sensing measurements on nanostructures were conducted in a custom-made high vacuum variable-temperature probe station equipped with a microheater. Prior to measurements, the sample was UV cleaned and annealed in a vacuum at $10^{-6}$ Torr for approximately 30 min at 350 °C. Pure oxygen ($10^{-4}$ Torr) was admitted to the vacuum chamber as a background gas mimicking oxidizing conditions of the ambient air. Computer-controlled solenoid valve pulses introduced reducing gases $H_2$ ($4 \cdot 10^{-4}$ Torr) and CO ($2 \cdot 10^{-4}$ Torr) into the chamber. The resultant changes in the source-drain current were measured as a function of time at a bias $V_{DS} = 6$ V. For photocatalytic measurements, a halogen lamp with a total light flux to the sample area of less than 0.2

W/cm$^2$ was used. Optical filters were added to block the ultraviolet and infrared spectrum regions of the light flux, eliminating artifacts related to the near UV and IR light components. Response of the nanostructures to reducing gases was measured as a function of the light illumination. A similar device fabricated from a SnO$_2$ individual nanowire was measured in the same conditions as the TiO$_2$ nanostructure for comparison.

**Results and Discussion**

The post-annealed crystalline structure of TiO$_2$ NWs as determined from the XRD pattern of Figure 1c is rutile. This polymorph has superior light scattering properties beneficial from the perspective of effective light harvesting, as compared with anatase.[21]

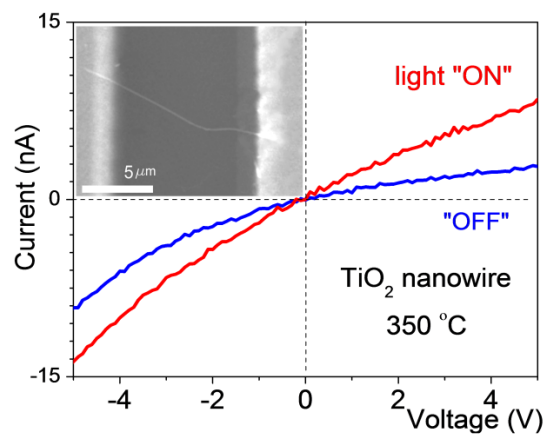

**Figure 2.** I-V curves of the TiO$_2$ nanowire device: The blue curve was recorded in the dark; the red curve shows electrical behavior under illumination with higher conductivity. The inset presents an SEM micrograph of the device.

The electrical behavior of titania NWs is shown in Figure 2. The blue and red IV curves were obtained in the absence and presence of illumination, respectively. Both of the curves appear to be non-linear, presumably due to the presence of a Schottky barrier at the nanowire-electrode contact. The IV curve recorded under illumination shows higher conductivity and deviates less from linearity than the one recorded in the dark. Several mechanisms can be responsible for this behavior. Firstly, visible-light assisted photoexcitation of electrons can be due to the existence of the filled impurity levels within the bandgap. These impurity levels become partially ionized and contribute new carriers in to conduction band under irradiation conditions. The electric field induced at the Schottky contact and initial band banding near nanowire surface both facilitate spatial separation of the hot electron-hole pairs and depresses the recombination rate, thus leading to a reduction of the Schottky barrier, band flattening and facilitated transport. Alternatively, energy of the visible light can be sufficient to induce photoinjection of electrons from the metallic electrode into the nanowire, which likewise increases overall conductivity.

The gas sensing properties of the fabricated $TiO_2$ NW microsensor were measured at 350 °C. At this elevated temperature the introduced background oxygen gas molecules easily dissociate and chemisorb onto the surface at the vacancy sites containing localized electrons and form $O^-$ ions, according to:[16]

$$O_{2(gas)} \rightarrow O_{2(ads)} \quad (1)$$

$$O_{2(ads)} + e^- \rightarrow O_{2(ads)}^- \quad (2)$$

$$O_{2(ads)}^- + e^- \rightarrow 2O_{(ads)}^- \quad (3)$$

Concurrently, energy bands bend upward decreasing the electron concentration in the near surface region and electron depletion layer is formed near the semiconductor surface.

A typical response of the TiO$_2$ NW device to admission of CO (2·10$^{-4}$ Torr) in to oxygen reach backround is shown in Figure 3. The lower blue curve was recorded with light "OFF", and the upper yellow ones were recorded under the visible light illumination. As can be seen, the overall background current as well as gas sensing response are elevated during illumination compared to the dark values similar to the IV curve data of Figure 2. In order to compare the gas responses ($\Delta I$) for both conditions more clearly, the background currents ($I_0$) were subtracted from the measured total current in the Figure 4. When CO was admitted to the chamber, a distinguishable decrease in nanowire resistance was observed within several seconds (Fig. 4). The NW's resistance recovered back to its original level upon interrupting CO inflow. The red and blue curves of Figure 4 show sensor's response to CO with and without light illumination, respectively. Approximately 10% higher response was achieved with the current intensity of visible light irradiation, as compared to that in the dark environment, despite the fact that the rise and recovery times are similar for both conditions. The TiO$_2$ NW response to H$_2$ (4·10$^{-4}$ Torr) exhibits similar result with approximately 25% higher response under illumination than in the dark, but again without a change in the response and recovery times (Fig 4b). These results demonstrate that the visible light photo activated TiO$_2$ NW is a good candidate material for photocatalysis and gas sensing.

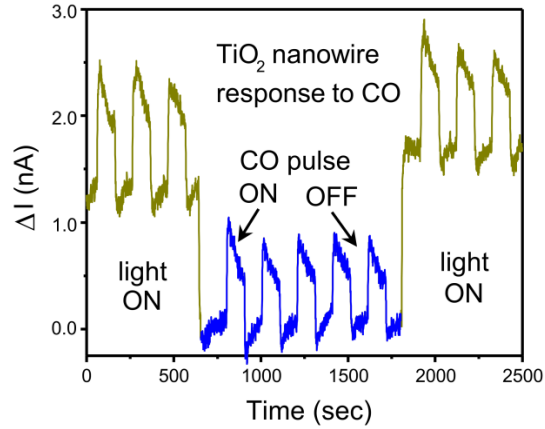

**Figure 3.** Response of individual TiO$_2$ NW to carbon monoxide (4·10$^{-4}$ Torr pulses) in the dark (blue curves) and under illumination (yellow curves) as measured at 350 °C.

There exist a few mechanisms that could explain these results. Herein, two scenarios are considered. First, the photo-assisted chemical sensing mechanism of a semiconductor material is discussed. The CO and H$_2$ molecules react with the adsorbed oxygen on the nanowire surface: [22]

$$O^- + CO_{(ads)} \rightarrow CO_{2(des)} + e^- \quad (5)$$

$$O^- + H_{2(ads)} \rightarrow H_2O_{(des)} + e^- \quad (6)$$

This process releases captured electrons back into the nanowire's surface leading to a reduction in the depletion layer width and a drop in the sensor's resistance. Carbon and/or Nitrogen doped TiO$_2$ has donor impurities levels in the bandgap, which can be photon excited by visible light irradiation to produce electron-hole pairs on the surface. The flattening of the bands promotes delivery of the photoexited electrons to the surface, where they can contribute to reduction of the preadsorbed oxygen species following Eq. (3). Thus, photoabsorption of the visible light increases the preadsorbed density of ionic oxygen (O⁻) at the TiO$_2$ nanowire surface, providing more active sites for further oxidation of CO and H$_2$.

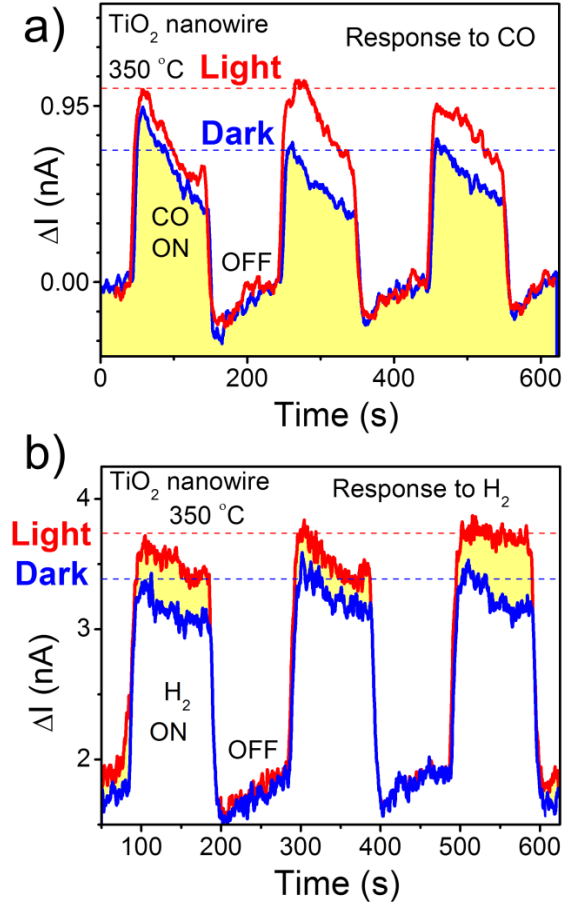

**Figure 4.** Response of individual TiO$_2$ NW to a) carbon monoxide and b) hydrogen ($4 \cdot 10^{-4}$ Torr pulses) in the dark and under illumination as measured at 350 °C.

Another plausible mechanism – Schottky diode – corroborates with previous studies on gas detection by metal-semiconductor structures (e.g. Au-carbon nanotube).[23-25] As the reducing gas molecules (H$_2$, CO) adsorb on the Au/TiO$_2$ interface, the work function of Au electrode is reduced, leading to a decrease in Schottky barrier (SB) for carrier injection to TiO$_2$. Consequently, the electrons that were either photogenerated on the dopant levels or photoexcited in the metal, can overcome the barrier, increasing the source-drain current:[24]

$$I_{DC} \sim T^2 e^{-\frac{q\phi_B}{kT}} \quad (7),$$

where $\phi_B$ is the Schottky barrier height, $k$ is the Boltzmann constant, $T$ is the absolute temperature, and $q$ is the elementary charge. In this case, the sensitivity of the SB modulation[24] is given by

$$S = \frac{\Delta R}{R_0} \approx e^{\frac{q\Delta\phi_B}{kT}} - 1 \quad (8)$$

Thus, because of the exponential dependence of the response on the changes in the SB height, a very high sensitivity can be achieved (as one shown in Fig. 4a and 4b). Therefore, the gas sensing response could be promoted by photoactivity on the $TiO_2$ NW surface in the visible light region.

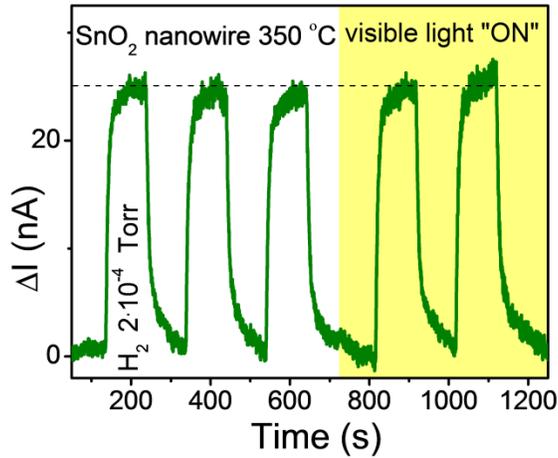

**Figure 5.** Response of individual $SnO_2$ NW to hydrogen ($4·10^{-4}$ Torr pulses) in the dark and under illumination (yellow region) as measured at 350 °C.

$SnO_2$[18] and ZnO could be a suitable alternative to $TiO_2$. A microsensor from an individual $SnO_2$ NW was fabricated in the same configuration and covered with contact micropads similar to those used for the $TiO_2$ NW device. The response of the $SnO_2$ NW to $H_2$ ($2 \cdot 10^{-4}$ Torr) was measured at 350 °C both in the dark and under visible light illumination. It appeared that its response is insensitive to illumination (Fig. 5), which gives less support to the proposed Schottky diode mechanism.

**Conclusion:**

Electrical measurements on individual $TiO_2$ single crystal nanowires imply that light induced electron-hole pair formation contribute to the increased electrical conductivity under visible light illumination. Furthermore, gas sensing (CO, $H_2$) measurements taken under both visible light irradiation and in the dark, indicate photo-activated chemical oxidization on the surface of $TiO_2$ nanowires. However, the precise mechanism of photo-assisted gas sensing is still unclear. It is speculated that $TiO_2$ single crystal nanowires were unintentionally doped with C and/or N during thermal growth process, which helped to expand their photoactivity to visible light. Future efforts will be focused on modifying the synthesis protocol to produce nanowires with higher sensing efficiency and controlled doping level.

**Acknowledgements:**

We would like to thank Dr. Y. Lilach for writing data acquisition LabView codes, Ms. K. Winspec for assistance with NW fabrication, and Mr. C. Watt for hardware setup.